\documentclass[a4paper,conference]{IEEEtran}

\usepackage{algorithm}
\usepackage{algpseudocode}
\usepackage{amsmath}
\usepackage{graphicx}
\usepackage{subcaption}
\usepackage{booktabs}
\usepackage{siunitx}
\algblock{Periodically}{EndPeriodically}
\algnewcommand\algorithmicperiodically{\textbf{periodically:}}
\algnewcommand\algorithmicendperiodically{} 

\title{ConflictSync: Bandwidth Efficient Synchronization of Divergent State}

 \author{
     \IEEEauthorblockN{
    Pedro Silva Gomes, Miguel Boaventura Rodrigues, Carlos Baquero
}
     \IEEEauthorblockA{
         MEIC\\
         Universidade do Porto \\
         Email: \texttt{pedromgomes29@gmail.com}
     }
     \IEEEauthorblockA{
         MEIC\\
         Universidade do Porto \\
         Email: \texttt{mbr@fe.up.pt}
     }
     \IEEEauthorblockA{
         Department of Informatics Engineering\\
         Universidade do Porto \\
         Email: \texttt{cbm@fe.up.pt}
     }
 }

\begin{document}

\maketitle

\begin{abstract}
State-based Conflict-free Replicated Data Types (CRDTs) are widely used in distributed systems to ensure high availability without coordination. However, their naive synchronization strategy—transmitting the full state—incurs high communication costs. Existing optimizations like $\delta$-CRDTs and $\Delta$-CRDTs reduce this overhead but rely on external metadata that must be garbage collected to prevent unbounded growth, at the cost of full state transmissions after network partitions.

This paper presents \textit{ConflictSync}, the first digest-driven synchronization algorithm for state-based CRDTs. We reduce synchronization to the set reconciliation of irredundant join decompositions and build on existing work in rateless set reconciliation. To support CRDTs, we generalize set reconciliation to variable-sized elements, and further introduce a novel combination of Bloom filters with rateless IBLTs to address inefficiencies at low similarity levels.

Our evaluation shows that ConflictSync reduces total data transfer by up to 18$\times$ compared to traditional state-based synchronization. Bloom filter prefiltering reduces overhead by up to 50\% compared to pure rateless reconciliation at 0\% similarity, while pure rateless reconciliation performs better above 93\% similarity. We characterize the trade-off between similarity level and Bloom filter size, identifying optimal configurations for different synchronization scenarios.

Although developed for CRDTs, ConflictSync applies to any synchronization problem where states can be decomposed into sets of constituent components, analogous to join decompositions, making it suitable for a wide range of distributed data models.
\end{abstract}
\begin{IEEEkeywords}
CRDTs, Synchronization, Replication, Set Reconciliation
\end{IEEEkeywords}

\section{Introduction}

Large-scale distributed systems often rely on replication to ensure fault tolerance, availability, and performance. However, replicating data introduces a trade-off between strong consistency and low latency \cite{cap}. Strong consistency models, such as linearizability, enforce a global ordering of operations but at the cost of availability and responsiveness. Eventual consistency offers a more relaxed model, allowing replicas to be updated concurrently without coordination. This leads to temporary divergence in replica states, which must be reconciled through synchronization mechanisms that preserve all updates.

Conflict-free Replicated Data Types (CRDTs) \cite{DBLP:conf/sss/ShapiroPBZ11} provide a principled approach to achieving eventual consistency. They ensure convergence without conflicts, even when updates occur concurrently. CRDTs are broadly classified into state-based and operation-based variants. In state-based CRDTs, replicas periodically exchange their full local states, merging them via a join operation. While this model tolerates unreliable channels, it can be inefficient when states are large or network bandwidth is limited.

To reduce the cost of synchronization, $\delta$-CRDTs \cite{DBLP:conf/netys/AlmeidaSB15} \cite{DBLP:conf/icde/EnesAB019} were introduced. These exchange small incremental states (\emph{deltas}) instead of full states and use acknowledgments to avoid redundant transmissions. However, they are not well suited to scenarios with high churn or infrequent communication. \(\Delta\)-CRDTs \cite{DBLP:conf/eurosys/LindeLP16} address some of these limitations by enabling replicas to compute minimal deltas based on causal metadata, such as vector clocks. Yet, they require additional metadata to be maintained and do not offer a general mechanism for deriving deltas across arbitrary CRDT designs.

Digest-driven synchronization offers a more efficient alternative by allowing replicas to exchange compact summaries of their states. One replica sends a digest, from which the peer can infer and transmit only the missing updates. This idea has been previously stated for CRDTs~\cite{digest_driven}, but no actual solution and implementation was provided.

We propose \emph{ConflictSync}, the first digest-driven synchronization algorithm. Our approach applies to any state-based CRDT that defines an irredundant join decomposition~\cite{DBLP:conf/icde/EnesAB019}. By decomposing CRDT states into sets of join-irreducible elements, we reduce synchronization to a set reconciliation problem. We leverage cryptographic hash functions, Bloom filters~\cite{DBLP:journals/cacm/Bloom70}, and Rateless Invertible Bloom Lookup Tables (Rateless IBLTs)~\cite{yang2024practical} to minimize communication costs.

While rateless set reconciliation has shown promise for fixed-size elements, applying it to variable-sized elements like join decompositions poses challenges. We address this by reconciling fixed-size digests of elements instead of the elements themselves, followed by fetching the full elements based on digest mismatches. Although rateless set reconciliation scales with the symmetric difference size, its high constant overhead makes it inefficient when replicas have low similarity. To address this, we devised a new synchronization strategy that integrates Bloom filter-based prefiltering with rateless IBLTs. 

Moreover, this new technique is not limited to the synchronization of CRDTs and also applies to the more general problem of synchronizing arbitrary sets of different sized elements. This result helps making rateless set reconciliation even more practical.


Our evaluation highlights the superior performance of ConflictSync across a wide range of similarity levels. At 0\% similarity, it reduces total transmission by 45\% compared to our generalization of rateless IBLTs for variable-sized elements. Even at 75\% similarity, ConflictSync still achieves a 36\% reduction in transmission. Notably, except for completely dissimilar states, ConflictSync consistently outperforms traditional state-driven synchronization, delivering transmission reductions of up to 18$\times$.

Summarizing, we make the following contributions:

\begin{enumerate}
    \item A generalization of Rateless Set Reconciliation for sets of variable sized elements, at the cost of additional communication steps.
    \item A new solution to generic set reconciliation, exhibiting very low metadata cost and minimal transmission of redundant data.
    \item An analysis of several different synchronization improvements and selection of the best approaches w.r.t. the data similarity profiles.
    \item An efficient solution for synchronization of state-based CRDTs, across different similarity levels.
\end{enumerate}

\section{Background}

This section provides an overview of the system model, state-based CRDTs, and synchronization techniques relevant to our approach.

\subsection{System Model}

Although our approach generalizes to multiple replicas in a connected topology via transitive pairwise synchronization, we focus on two replicas, A and B, referred to simply as replicas. Communication between replicas occurs over a reliable, bidirectional, asynchronous FIFO channel, which may break and later be re-established. This can result in end-to-end retransmission of unacknowledged data, similar to a TCP connection that fails and resumes in a new session.

We assume a crash-recovery model, where replicas may fail at any point and later recover with a previously valid state. Our algorithms leverage CRDTs to ensure correct state convergence despite retransmissions, message reordering, or dropped messages. Furthermore, during synchronization, replicas must not accept any operations that would alter their state\footnote{To preserve availability, the state used for synchronization can be forked from the state accepting user operations and later re-merged after synchronization.}.

\subsection{CRDTs}

CRDTs are data types that guarantee convergence without coordination in eventually consistent systems. They allow multi-master replication where any replica can accept updates, even if the network is unavailable. When network communication is available CRDTs ensure that all replicas eventually converge to the same state without exposing conflicts to the user. CRDTs come in two main variants: \emph{state-based} and \emph{operation-based}.  

In \emph{state-based} CRDTs, the set of possible states forms a \emph{join-semilattice} \cite{lattices&order}, an order-theoretic structure that enables a well-defined merging operation. Given two states, \(s\) and \(s'\), their merge is defined as the \emph{least upper bound}, \(s \sqcup s'\). State updates occur through \emph{mutators}, which must be \emph{inflationary}, meaning that applying a mutation \(m(s)\) results in a state satisfying \(s \sqsubseteq m(s)\).  Where \(\sqsubseteq\) is an order relation on states.

If a data type ensures that both its mutators and merge operations satisfy these properties, and if messages are eventually delivered, then the data type is \emph{strongly eventually consistent} by construction \cite{DBLP:conf/sss/ShapiroPBZ11}.

\begin{figure}
\begin{eqnarray*}
\text{GSet}\langle E \rangle & = & \mathcal{P}(E) \\
\bot & = & \emptyset \\
\text{add}(e, s) & = & s \cup \{e\} \\
\text{value}(s) & = & s \\
s \sqcup s' & = & s \cup s'
\end{eqnarray*}
    \caption{State-based Grow-Only Set.}
    \label{fig:state-based-gset}
\end{figure}

In Figure \ref{fig:state-based-gset}, we define a simple state-based CRDT: the Grow-Only Set (GSet). This data type represents a set that supports only the addition of elements. The add operation inserts an element into the set and returns the updated set, while the join operation computes the union of two sets. Although this is a simple example, the techniques and results presented in this paper apply to any state-based CRDT, including Add-Wins Sets and map CRDTs that embed other CRDTs as values~\cite{approaches_crdts}.

\subsection{Join Decompositions}\label{sec:join-decomp}

We have seen that \emph{state-based} CRDT states correspond to an element of an appropriate \emph{join-semilattice} and that any two elements can be joined by the corresponding join $\sqcup$. In order to synchronize CRDT replicas, without transferring the whole state, it is relevant to allow decomposing a lattice state into smaller lattice states from the same lattice.   

The \emph{irredundant join decomposition} \cite{DBLP:conf/icde/EnesAB019} of a state \(s\) represents its decomposition into a set of smaller states that, when joined, reconstruct \(s\). Each state in the decomposition is \emph{join-irreducible}, meaning it cannot be further decomposed. Moreover, the decomposition is \emph{irredundant}, ensuring that no state in the set is unnecessary: removing any element would prevent the remaining states from merging back into \(s\).  

For example\footnote{For a complete catalog of decompositions refer to \cite{DBLP:conf/icde/EnesAB019}.}, the irredundant join decomposition of a GSet state \(s\) is:  
\[
\Downarrow s = \{\{e\} \mid e \in s\}
\]

If \( s = \{a, b, c\} \), then its decomposition is \( \Downarrow s = \{\{a\}, \{b\}, \{c\}\} \). Notice that  
\[
\{a\} \sqcup \{b\} \sqcup \{c\} = \{a,b,c\} = s,
\]
and neither \(\{a\}\), \(\{b\}\), nor \(\{c\}\) can be removed without altering the resulting state upon merging. Furthermore, none of these elements can be further decomposed.

Given a unique irredundant join decomposition, the minimum delta (or "difference") between two states \(a\) and \(b\) is:
\[
\Delta(a, b) = \bigsqcup \{ y \in \Downarrow a \mid y \not\in b \}
\]
It represents the smallest state that, when joined with \(b\), reconstructs \(a \sqcup b\).

\subsection{Synchronization Algorithms}
\subsubsection{RSync}  
The \emph{RSync} algorithm \cite{tridgell1996rsync} is widely used to efficiently synchronize a file on a local machine with its corresponding version on a remote machine. The synchronization direction must be defined. A replica is declared primary, having the most up-to-date information, while the other replica is secondary and must be synchronized. The core idea is to partition the local file into blocks, exchange the signatures of those blocks, and transfer only the non-matching blocks—i.e., those whose signatures differ.  

A straightforward approach would be to divide the file into fixed-size blocks. However, this method fails to detect matches that do not align with block boundaries. For instance, if a single character is inserted at the beginning of the file, all block boundaries shift, preventing any matches and requiring the transmission of the entire file.  

A naive solution would be to compute signatures at every possible block boundary on the receiver side. However, cryptographically secure signatures with negligible collision probability are computationally expensive, making this approach infeasible. To address this, RSync employs two signatures: weak and strong signatures. Weak signatures are computationally efficient but have a non-negligible collision probability. They serve as a preliminary filter to limit the number of strong signature computations. Strong signatures, in contrast, have an extremely low collision probability and are used to reliably detect matching blocks, thereby eliminating the need to transmit them over the network.

\subsubsection{Bloom Filters}\label{section:bloom}
A Bloom filter \cite{DBLP:journals/cacm/Bloom70} is a space-efficient probabilistic data structure that represents a set and supports membership queries. It consists of an array of \( m \) bits, all initially set to 0, and utilizes \( k \) hash functions that map each element to \( k \) positions within the array.  

To insert an element, the Bloom filter applies the \( k \) hash functions to determine \( k \) positions and sets the corresponding bits to 1. To check whether an element is in the set, the same \( k \) hash functions are applied, and the filter verifies whether all corresponding bits are set to 1. Since the hash functions are deterministic, an inserted element will always map to the same positions, ensuring that previously set bits remain unchanged in future queries.  

If at least one of the \( k \) bits is 0, it is guaranteed that the element is not in the set. However, multiple insertions can independently set all \(k\) bits corresponding to an element that was never inserted, causing the filter to incorrectly indicate its presence. This results in false positives, where the filter incorrectly indicates the presence of an element. However, Bloom filters never produce false negatives, meaning they will never mistakenly indicate that a present element is absent. Standard Bloom filters are grow-only and do not support element removal.

\subsubsection{Rateless Set Reconciliation}  

Invertible Bloom Lookup Tables (IBLTs) \cite{eppstein2010straggler} \cite{goodrich2011invertible} extend Bloom filters by allowing element removal and enabling the retrieval of the stored set elements with high probability while requiring space proportional to the number of elements present at the time of listing, even if the set previously contained a significantly larger number of elements. This is achieved by storing additional information in each cell instead of a single bit, as in Bloom filters. Specifically, each cell maintains the following fields:  

\begin{itemize}  
    \item \textbf{idSum:} The XOR sum of all elements mapped to the cell.  
    \item \textbf{hashSum:} The XOR sum of the hash values of all elements mapped to the cell.  
\end{itemize}  

To reconstruct the set elements, the decoder executes a recursive process known as \emph{peeling}. It begins by identifying a \emph{pure cell}, which is a cell where the hash of \texttt{idSum} matches \texttt{hashSum}, indicating that \texttt{idSum} corresponds to one of the original elements inserted into the cell. Once a pure cell is found, the corresponding element is recovered and removed from all other cells to which it was mapped. This process may reveal new pure cells, allowing the procedure to continue iteratively until no pure cells remain. Decoding fails if the process terminates before all original elements are recovered.  

Given two sets \( S_A \) and \( S_B \), their respective IBLTs, \( \mathit{IBLT}_A \) and \( \mathit{IBLT}_B \), can be combined to compute \( \mathit{IBLT}_{S_A \Delta S_B} \), the IBLT of the symmetric difference \( S_A \Delta S_B \), where \( S_A \Delta S_B \doteq (S_A \cup S_B) \setminus (S_A \cap S_B) \). This is achieved by applying the XOR operation cell-wise between \( \mathit{IBLT}_A \) and \( \mathit{IBLT}_B \): for each cell, the `idSum` and `hashSum` fields are XORed as \( \text{idSum}_A \oplus \text{idSum}_B \) and \( \text{hashSum}_A \oplus \text{hashSum}_B \). The XOR operation cancels out elements that appear in both sets, effectively removing common elements from the resulting IBLT. For example, since \( x \oplus x = 0 \), any element shared by both \( S_A \) and \( S_B \) will be eliminated from the symmetric difference.

This approach enables set reconciliation \cite{eppstein2011s}: Node \( A \) sends \( \mathit{IBLT}_A \) to Node \( B \), which combines it with its own \( \mathit{IBLT}_B \) to compute \( \mathit{IBLT}_{S_A \Delta S_B} \). If the IBLT is appropriately sized (i.e., proportional to the number of differing elements), all elements in the set difference can be recovered. However, a fundamental challenge lies in determining the required number of cells, as node \( A \) does not know the exact number of differing elements. Existing algorithms estimate the set difference size, but overestimation leads to unnecessary communication overhead, while underestimation increases the risk of decoding failure, requiring protocol restarts.  

With conventional IBLTs, if decoding fails with a given number of cells \( m \) and a larger size \( m + k \) is needed, both nodes must rebuild and retransmit the entire IBLT, since the mapping of elements to cells changes.  

\emph{Rateless Set Reconciliation} \cite{yang2024practical} addresses this issue by allowing node \( A \) to stream an unbounded sequence of coded symbols (IBLT cells) to node \( B \), which generates its own corresponding stream. A prefix of length \( m \) from these streams enables reconciliation of \( O(m) \) set differences. If decoding fails for a given prefix size \( m \), and a larger size \( m + k \) is required, rateless IBLTs allow for incremental transmission: only the additional \( k \) symbols need to be generated and sent, as both the first \( m \) cells and the extended \( m + k \) cells remain consistent prefixes of the same infinite coded sequence.  

In practice, node \( A \) continuously generates and transmits coded symbols to node \( B \) until \( B \) confirms having received a sufficient number of symbols to recover all elements in the set difference. Simulations indicate that the expected number of coded symbols required to synchronize \( d \) differences ranges from \( 1.72d \) for small set differences to approximately \( 1.35d \) as \( d \) increases \cite{yang2024practical}.

\section{ConflictSync Algorithms}\label{section:algos}
This section introduces a set of algorithms designed for the efficient synchronization of state-based CRDTs. While the focus is on CRDTs, the underlying techniques are broadly applicable to any system that involves the synchronization of arbitrary sets, or large states that can be decomposed into sets of smaller components - similar to irredundant join decompositions in the context of state-based CRDTs.
Adapting to general sets is easily obtained by ignoring the decomposition operator $\Downarrow$ and assuming states are already sets.

The first four algorithms were initially introduced in 
the second author's MSc Thesis \cite{crdt_infrastructure},
while the following algorithms improve upon them by integrating techniques from the literature on set reconciliation.

\subsection{State-Driven Synchronization}\label{section:state-driven-sync}

\begin{algorithm}
\caption{State-driven synchronization}
\label{alg:state-driven-sync}
\begin{algorithmic}[1]
    \State \textbf{durable State:}
    \State \hspace{1em} $X_A, X_B \in S$
    
    \Periodically
    \State $\text{send}_{A,B}(\text{Sync}, X_A)$ 
    \EndPeriodically
    
    \Procedure{OnReceive$_{A,B}$}{$\text{Sync}, X_A$}
        \State $d = \Delta(X_B, X_A)$ 
        \State $X_B \gets X_B \sqcup X_A$ 
        \State $\text{send}_{B,A}(\text{MissingState}, d)$
    \EndProcedure

    \Procedure{OnReceive$_{B,A}$}{$\text{MissingState}, d$}
        \State $X_A \gets X_A \sqcup d$ 
    \EndProcedure
\end{algorithmic}
\end{algorithm}

In this approach, replica \( A \) sends its full state \( X_A \) to replica \( B \). Upon receiving \( X_A \), replica \( B \) computes the delta \( d = \Delta(X_B, X_A) \), where \(\Delta\) is the operator defined in Subsection \ref{sec:join-decomp}. This delta represents the portion of \( X_B \) that strictly inflates \( X_A \). \( B \) then sends \( d \) back to \( A \), allowing \( A \) to update its state. This method ensures that replica \( B \) only sends the minimal necessary state to \( A \). However, this synchronization is inefficient when \( X_A \) and \( X_B \) are similar, as \( A \) still transmits its full state, which is not optimal. The synchronization protocol, initiated by replica \( A \), is formally defined in Algorithm~\ref{alg:state-driven-sync}.
This algorithm serves as the baseline for evaluating other synchronization strategies in this work.

\subsection{Bucketing}\label{sec:bucketing}
Bucketing aggregates join decompositions into buckets, enabling synchronization at a more granular level. Each bucket forms a CRDT state by merging its assigned decompositions, and a digest is computed for each bucket. Replicas exchange these digests and synchronize only buckets with mismatched digests, following the state-driven synchronization process as explained in Section \ref{section:state-driven-sync}.

Join decompositions are assigned to buckets using a hash partitioning strategy. For each decomposition, a hash function is applied to compute a hash value. The decomposition is then placed into a bucket determined by taking the modulo of this hash value with respect to the total number of buckets. To ensure consistent bucket digests across replicas, each bucket's contents must be mapped deterministically. This is achieved by sorting join decompositions within a bucket by their hash values and computing the digest as the hash of the concatenated sorted hashes.

A key drawback is that state transfer may exceed the actual differences in join decompositions. Only fully identical buckets can be skipped, while similar but nonidentical buckets must be fully exchanged. In a worst-case scenario, if differences are evenly distributed across buckets, the total data transfer may be significantly larger than the actual difference. For example, consider 10 buckets containing a total of 100 join decompositions, where 90 are already shared with the other replica and 10 are newly added and unknown to it. If these 10 new decompositions are distributed across all 10 buckets, each bucket will appear out-of-sync, triggering synchronization for all buckets. As a result, all 100 decompositions are transferred instead of just the 10 that differ, incurring a \(10\times\) overhead.

\subsection{Bloom-based}
The synchronization of state-based CRDTs can be framed as a set-reconciliation problem, specifically, the synchronization of the sets of irredundant join decompositions, \(\Downarrow X_A\) and \(\Downarrow X_B\). Bloom filters, as described in Subsection~\ref{section:bloom}, are a space-efficient probabilistic data structure that can approximately solve the set reconciliation problem.

The algorithm works as follows: Replica \(A\) inserts all of its join decompositions into a Bloom filter, \(BF_A\), and transmits it to replica B. Upon receiving \(BF_A\), B checks the presence of its own join decompositions, partitioning \(X_B\) into two sets:
\begin{itemize}
    \item \textbf{Exclusive join decompositions}, \( X_B^{\text{excl}} \), which consists of join decompositions that are definitely not in \( X_A \) (true negatives).
    \item \textbf{Potentially common join decompositions}, \( X_B^{\text{com}} \), including join decompositions that may already exist in \( X_A \) (true positives and false positives).
\end{itemize}

Replica \(B\) then constructs a Bloom filter, \(\textit{BF}_B\), from \(X_B^{\text{com}}\) and sends it to A, along with all join decompositions in \( X_B^{\text{excl}} \). Upon receiving \(\textit{BF}_B\), \(A\) similarly partitions \(X_A\) into \( X_A^{\text{excl}} \) and \( X_A^{\text{com}} \) and transmits \( X_A^{\text{excl}} \) back to \(B\).

While Bloom filters are highly space-efficient, they introduce false positives, meaning that after synchronization, replicas \(A\) and \(B\) may not necessarily be fully consistent. Bloom filters can be used as a preliminary step to significantly increase the similarity between two sets. This allows a subsequent synchronization algorithm - one that guarantees convergence - to operate over sets that are already much closer in content. 

This approach is particularly advantageous for synchronization algorithms that perform very well when the symmetric difference between sets is small, but do not adapt well as the difference grows.

This strategy - using Bloom filters to first greatly increase the similarity between sets, followed by a reconciliation phase limited to correcting false positives - is employed in the algorithms described in Subsections~\ref{section:bloom+bucketing}, \ref{section:bloom+rateless}, and \ref{section:bloom+bucketing+rateless}.

\subsection{Bloom-based + Bucketing}\label{section:bloom+bucketing}
A key challenge with the bucketing approach is the potential for significant redundancy in the transmission of similar but non-identical buckets. In contrast, the main limitation of the Bloom filter approach is the lack of convergence. These two methods can be effectively combined: Bloom filters are used to partition \(X_A\) and \(X_B\) into exclusive and potentially common join decompositions, while bucketing is applied to synchronize the false positives remaining in \(X_A^{\text{com}}\) and \(X_B^{\text{com}}\). 

Assuming an adequately sized Bloom filter, the proportion of false positives is small, ensuring that \(X_A^{\text{com}}\) and \(X_B^{\text{com}}\) are highly similar. This minimizes the occurrence of similar but non-identical buckets, thereby mitigating the redundancy issue inherent in bucketing. Furthermore, the use of bucketing not only detects false positives but also ensures convergence, addressing the limitation of Bloom filters alone. This combined approach is analogous to the RSync protocol, where Bloom filters act as the weak checksum and bucketing serves as the strong checksum.

A downside of this approach is increased latency. Both the Bloom-based and bucketing methods require 3 message exchanges, resulting in a latency of 1.5 RTT. In contrast, the combined approach incurs a latency of 2 RTT.

\subsection{Rateless Set Reconciliation}

As explained previously, the problem of synchronizing state-based CRDTs can be reduced to synchronizing the corresponding set of join decompositions. Rateless set reconciliation provides a compelling solution to this problem; however, our setting deviates slightly from the classical formulation: join decompositions are variable-sized, whereas traditional set reconciliation methods assume fixed-size elements of length \(k\). A simple workaround is to reconcile fixed-length cryptographic hashes of variable-sized elements instead. This idea - generalizing set reconciliation to variable-sized elements by hashing - has been mentioned in prior work (see, e.g., \cite{practical_set_reconciliation}). Our work presents a concrete algorithm that applies this idea and provides an empirical evaluation in the context of CRDT synchronization.

The full protocol, initiated by replica \( A \), is formally defined in Algorithm~\ref{alg:rateless-sync}.

\begin{algorithm}
\caption{Rateless Join Decomposition Set Reconciliation}
\label{alg:rateless-sync}
\begin{algorithmic}[1]

    \State \textbf{durable state:}
    \State \hspace{1em} $X_A, X_B \in S$ \Comment{CRDT states at replicas A and B}
    
    \Periodically
        \State  $H_A \gets \{ \text{hash}(\text{jd}) \mid \text{jd} \in \Downarrow X_A \}$  
        \State $i \gets 0$  
    
        \While{B has not signaled completion}
            \State ${\text{IBLT}_{H_A}}[i] \gets \text{generateCodedSymbol}(H_A, i)$ 
            \State $\text{send}_{A,B}(\text{SymStream}, i, {\text{IBLT}_{H_A}}[i])$  
            \State $i \gets i + 1$  
        \EndWhile
    \EndPeriodically

    \Procedure{OnReceive$_{A,B}$}{$\text{SymStream}, i, IBLT_{H_A}[i]$}
        \If{$i==0$}
            \State  $H_B \gets \{ \text{hash}(jd) \mid jd \in \Downarrow X_B \}$  
        \EndIf
        \State ${\text{IBLT}_{H_B}}[i] \gets \text{generateCodedSymbol}(H_B, i)$ 
        \State ${\text{IBLT}_{{H_B}\Delta{H_A}}}[i] \gets {\text{IBLT}_{H_B}}[i] \oplus {\text{IBLT}_{H_A}}[i]$
        
        \If{${\text{IBLT}_{{H_B}\Delta{H_A}}}$ is decodable}
            \State $(H_{A \setminus B}, H_{B \setminus A}) \gets \text{decode}({\text{IBLT}_{{H_B}\Delta{H_A}}})$  
            \State $X_{B \setminus A} \gets \bigsqcup\{ \text{jd} \mid \text{jd} \in \Downarrow X_B, \text{hash}(\text{jd}) \in H_{B \setminus A} \}$ 
            \State $\text{send}_{B,A}(\text{EOS}, H_{A \setminus B}, X_{B \setminus A})$  
        \EndIf  
    \EndProcedure

    \Procedure{OnReceive$_{B,A}$}{$\text{EOS}, H_{A \setminus B}, X_{B \setminus A}$}
        \State $X_{A \setminus B} \gets \bigsqcup\{ \text{jd} \mid \text{jd} \in \Downarrow X_A, \text{hash}(\text{jd}) \in H_{A \setminus B} \}$
        \State $X_A \gets X_A \sqcup X_{B \setminus A}$
        \State $\text{send}_{A,B}(\text{MissingState}, X_{A \setminus B})$
    \EndProcedure

    \Procedure{OnReceive$_{A,B}$}{$\text{MissingState}, X_{A \setminus B}$}
        \State $X_B \gets X_B \sqcup X_{A \setminus B}$
    \EndProcedure

\end{algorithmic}
\end{algorithm}

Replica \( A \) streams coded IBLT symbols to replica \( B \) via \texttt{SymStream} messages until \( B \) has received a sufficient number of symbols to decode \( H_{B \Delta A} \), the symmetric difference between the sets of hashes \( H_A \) and \( H_B \). These sets represent the join decomposition hashes stored at replicas \( A \) and \( B \), respectively. For details on the \texttt{generateCodedSymbol} function, the \(\oplus\) operator, the decodability check, and the decoding process, refer to the original work on Rateless Set Reconciliation~\cite{yang2024practical}.

Once decoding is successful, replica \( B \) identifies the set of hashes missing at \( A \), denoted \( H_{B \setminus A} \), retrieves the corresponding join decompositions, and aggregates them into \( X_{B \setminus A} \). It then prepares to send these decompositions to \( A \).

Although \( B \) also learns the hashes it is missing, \( H_{A \setminus B} \), it still lacks the actual data. To retrieve the missing decompositions, it sends \( H_{A \setminus B} \) and \( X_{B \setminus A} \) to \( A \) in a \texttt{EOS} (End of Stream) message. This enables \( A \) to identify the missing data and respond accordingly.

Upon receipt, replica \( A \) extracts the relevant decompositions into \( X_{A \setminus B} \) and integrates \( X_{B \setminus A} \) into its local state. It then sends \( X_{A \setminus B} \) back to \( B \) in a \texttt{MissingState} message, allowing \( B \) to complete the synchronization process by merging the data into its own state.

\subsection{Bloom-based + Rateless Set Reconciliation}\label{section:bloom+rateless}

\begin{algorithm}
\caption{Rateless Join Decomposition Set Reconciliation with Bloom Filters}
\label{alg:bloom-rateless-sync}
\begin{algorithmic}[1]
    \State \textbf{durable state:}
    \State \hspace{1em} $X_A, X_B \in S$ \Comment{CRDT states at replicas A and B}

    \Periodically
        \State $\text{BF}_A \gets  \text{buildFilter}(\text{X}_A)$
        \State $\text{send}_{A,B}(\text{Bloom}, \text{BF}_A)$
    \EndPeriodically
        
    \Procedure{OnReceive$_{A,B}$}{$\text{Bloom}, \text{BF}_A$}
        \State $X_B^{\text{com}} \gets \bigsqcup \{ y \mid y \in \Downarrow X_B \land y \in \text{BF}_A \}$
        \State $X_B^{\text{excl}} \gets \bigsqcup \{ y \mid y \in \Downarrow X_B \land y \notin \text{BF}_A \}$ 
        \State $\text{BF}_B \gets  \text{buildFilter}(\text{X}_B^{\text{com}})$

        \State $\text{send}_{A,B}(\text{InitStream}, BF_B, X_B^{\text{excl}})$

        \State $H_B\gets \{ \text{hash}(\text{jd}) \mid \text{jd} \in X_B^{\text{com}} \}$  

        \While{A has not signaled completion}
            \State ${IBLT_{H_B}}[i] \gets \text{generateCodedSymbol}(H_B, i)$  
            \State $\text{send}_{B,A}(\text{SymStream}, i, {\text{IBLT}_{H_B}}[i])$  
            \State $i \gets i + 1$  
        \EndWhile

    \EndProcedure

    \Procedure{OnReceive$_{B,A}$}{$\text{InitStream},  \text{BF}_B, X_B^{\text{excl}}$}
        \State $X_A^{\text{com}} \gets \bigsqcup \{ y \mid y \in \Downarrow X_A \land y \in \text{BF}_B \}$
        \State $X_A^{\text{excl}} \gets \bigsqcup \{ y \mid y \in \Downarrow X_A \land y \notin \text{BF}_B \}$ 
        \State $H_A\gets \{ \text{hash}(\text{jd}) \mid \text{jd} \in X_A^{\text{com}} \}$
        \State $X_A \gets X_A \sqcup X_B^{\text{excl}}$

    \EndProcedure

    \Procedure{OnReceive$_{B,A}$}{$\text{SymStream}, i, \text{IBLT}_{H_B}[i]$}
        \State ${\text{IBLT}_{H_A}}[i] \gets \text{generateCodedSymbol}(H_A, i)$ 
        \State ${\text{IBLT}_{{H_A}\Delta{H_B}}}[i] \gets {\text{IBLT}_{H_A}}[i] \oplus {\text{IBLT}_{H_B}}[i]$
        
        \If{${\text{IBLT}_{{H_A}\Delta{H_B}}}$ is decodable}
            \State $(H_{A\setminus B}, H_{B \setminus A}) \gets \text{decode}({\text{IBLT}_{{H_A}\Delta{H_B}}})$  
            \State $X_A^{\text{FP}} \gets \bigsqcup\{ \text{jd} \mid \text{jd} \in \Downarrow X_A, \text{hash}(\text{jd}) \in H_{A\setminus B} \}$
            \State $\text{send}_{A,B}(\text{EOS}, H_{B\setminus A}, X_A^{\text{FP}}, X_A^{\text{excl}})$  
        \EndIf  
    \EndProcedure

    \Procedure{OnReceive$_{A,B}$}{$\text{EOS}, H_{B\setminus A}, X_A^{\text{FP}}, X_A^{\text{excl}}$}
        \State $X_B^{\text{FP}} \gets \bigsqcup\{ jd \mid jd \in \Downarrow X_B, \text{hash}(jd) \in H_{B\setminus A} \}$
        \State $X_B \gets X_B \sqcup X_A^{\text{FP}}$
        \State $X_B \gets X_B \sqcup X_A^{\text{excl}}$
        \State $\text{send}_{B,A}(\text{MissingFP}, X_B^{\text{FP}})$
    \EndProcedure

    \Procedure{OnReceive$_{B,A}$}{$\text{MissingFP}, X_B^{\text{FP}}$}
        \State $X_A \gets X_A \sqcup X_B^{\text{FP}}$
    \EndProcedure

\end{algorithmic}
\end{algorithm}

Rateless set reconciliation allows the synchronization of sets with fixed-size elements, with the communication cost being proportional to the size of the symmetric difference. However, the overhead associated with this method can be significant. The expected number of coded symbols required typically ranges from $1.35d$ to $1.72d$, and while the \texttt{idSum} field contributes to the actual data transmission, the \texttt{hashSum} and \texttt{count} fields introduce substantial overhead, particularly when the \texttt{idSum} field is small.

Furthermore, because the synchronization occurs at the level of hashes of join decompositions rather than the decompositions themselves, additional communication is required for replica \( B \) to request \( X_{A \setminus B} \) once it has determined \( H_{A \setminus B} \).

In contrast, Bloom filters offer a more efficient alternative by minimizing constant overhead. The size of a Bloom filter is proportional to the number of elements being transmitted, with an accuracy of 97.8\% achievable using eight bits per element and five hash functions \cite{approximate_set_reconciliation}. For example, a Bloom filter for 10,000 elements requires only 80,000 bits (10 KB).

As demonstrated in the evaluation section, rateless set reconciliation becomes less efficient when the dissimilarity between replicas is high. To mitigate this, we combine rateless reconciliation with Bloom filters. Replica \( A \) begins the process by constructing a Bloom filter over its join decompositions and sending it to replica \( B \) in a \texttt{Bloom} message. Upon receiving the message, replica \( B \) partitions its local decompositions into two subsets: \( X_B^{\text{com}} \), which matches the filter, and \( X_B^{\text{excl}} \), which does not. It then replies with an \texttt{InitStream} message containing a Bloom filter for \( X_B^{\text{com}} \) and the set \( X_B^{\text{excl}} \).

Replica \( A \) uses the second Bloom filter to extract \( X_A^{\text{com}} \) and \( X_A^{\text{excl}} \). The exclusive elements \( X_B^{\text{excl}} \) are merged immediately. Rateless set reconciliation then proceeds over the hash sets \( X_A^{\text{com}} \) and \( X_B^{\text{com}} \), as outlined in the previous subsection.

The \texttt{EOS} message sent by replica \( A \) after decoding contains three fields: \( H_{B \setminus A} \), which consists of the hashes of the join decompositions in \( X_B^{\text{FP}} \), along with \( X_A^{\text{FP}} \) and \( X_A^{\text{excl}} \). Replica \( B \) merges \( X_A^{\text{FP}} \) and \( X_A^{\text{excl}} \) into its state, computes \( X_B^{\text{FP}} \) using \( H_{B \setminus A} \), and sends a \texttt{MissingFP} message back to replica \( A \), containing the computed \( X_B^{\text{FP}} \).  Upon receipt, replica \(A\)
A finalizes the protocol by merging \( X_B^{\text{FP}} \) into its own state.

The synchronization protocol is formally defined in Algorithm~\ref{alg:bloom-rateless-sync}.

This approach shares a key drawback with $\textit{Bloom-based + Bucketing}$: a latency of 2 RTT. However, unlike $\textit{Bloom-based + Bucketing}$, which requires exchanging hashes for all buckets, the communication overhead (excluding the bloom filters) is proportional only to the number of differing hashes. Since only false positives remain to be synchronized, this overhead is significantly lower.

Interestingly, a similar idea of combining Bloom filters and (non-rateless) invertible Bloom lookup tables (IBLTs) has been previously explored in Graphene \cite{graphene}. In our work, rateless set reconciliation was initially used to synchronize the digests of the join decompositions, but it performed poorly when the similarity between replicas was low. This observation led us to reuse the idea, previously applied in the $\textit{Bloom-based + Bucketing}$ algorithm, of employing Bloom filters to increase the similarity between sets before applying a second synchronization algorithm.

Unlike Graphene, which relies on standard IBLTs and provides only probabilistic convergence guarantees, our approach leverages rateless set reconciliation. This distinction is critical because it removes the need to predetermine the size of the IBLT, ensuring convergence without relying on probabilistic bounds. We explore this distinction further in the Related Work section.

\subsection{Bucketing + Rateless Set Reconciliation}

In the bucketing approach (Section~\ref{sec:bucketing}), one replica sends the digests of all buckets, incurring a communication cost proportional to the total number of buckets. This can be optimized by using rateless set reconciliation to compute the symmetric difference over bucket digests, reducing communication to depend only on the number of differing buckets.

However, this introduces overhead. The symmetric difference includes two hashes per differing bucket, doubling the cost compared to standard bucketing when all buckets differ. Furthermore, as mentioned in Section \ref{section:bloom+rateless}, the rateless approach incurs constant overhead per symbol (due to \texttt{hashSum} and \texttt{count}) and requires $1.35d$ to $1.72d$ symbols on average. Still, when most buckets match, this method significantly reduces the data exchanged while preserving convergence guarantees.

\subsection{Bloom-based + Bucketing + Rateless Set Reconciliation}\label{section:bloom+bucketing+rateless}

This approach integrates Bloom filters, bucketing, and rateless set reconciliation, techniques that have been previously detailed. Bloom filters are employed to partition the sets of join decompositions into exclusive and potentially common elements. These common decompositions are then grouped into buckets, and rateless set reconciliation is used to identify only the mismatched buckets. 

\section{Evaluation}\label{sec:evaluation}

\subsection{Experimental Setup}

We implemented and benchmarked all algorithms described in Section \ref{section:algos}, excluding the Bloom-based approach, as it is probabilistic and does not guarantee synchronization.

To benchmark the algorithms, we used a simulator to replicate real-world conditions. Although the experiments were executed on a single machine, we carefully modeled the data transmission that would occur in a real network. For each algorithm, we measured three key metrics: (i) the metadata sent (e.g., Bloom filters, bucket hashes), (ii) the redundancy sent (i.e., join decompositions that were unnecessary to transmit because they were already present at both replicas), and (iii) the total transmitted bytes. This approach allowed us to assess the communication overhead and the efficiency of each method in terms of data transfer.

For algorithms with tunable parameters, we explored various configurations to examine their impact on performance. Specifically, for methods utilizing bucketing, we varied the load factor \( f_{ld} \), which determines the number of buckets for a given state \( X \), as defined by the following formula:

\[
B(X) = \left\lfloor |\Downarrow X| \cdot f_{ld} \right\rfloor, \quad \text{where } f_{ld} > 0.
\]

The expected number of join decompositions per bucket is \( \frac{1}{f_{ld}} \). For example, when \( f_{ld} = 0.2 \), each bucket is expected to contain five join decompositions, whereas for \( f_{ld} = 5 \), the expected number per bucket is 0.2. A lower \( f_{ld} \) results in fewer exchanged hashes but decreases the likelihood of hash matches, leading to increased redundant state transfer. Conversely, a higher \( f_{ld} \) increases the number of exchanged hashes, enhancing the chances of matching and thus reducing redundant state transmission.

It is important to note that both replicas must use the same number of buckets. Therefore, while \( f_{ld} \) may not be identical across replicas, the replica initiating the synchronization procedure determines the number of buckets based on its chosen \( f_{ld} \), and the other replica simply uses the same number.

For methods using Bloom filters, we varied the probability of false positives \( \epsilon \) by adjusting the size of the Bloom filter. Unlike in bucketing approaches, the Bloom filter size can differ between replicas, so the same \( \epsilon \) value is maintained at both ends.

We used GSets in our analyses. Each experiment generates two replica states, \( X_A \) and \( X_B \), with a given similarity \(s\). The similarity \( s \) is computed using the Jaccard index between the irredundant join decompositions of \( X_A \) and \( X_B \). Thus, \( s \in [0,1] \), where \( s = 0 \) indicates disjoint sets:

\[
s = J(\Downarrow X_A, \Downarrow X_B)
\]

For the GSet experiments, we generated sets with 100,000 distinct items, where each item is a string with a size uniformly distributed over the interval \([5,80]\). These items were selected based on the similarity metric, and the cardinality of the sets remained consistent across replicas. The use of a range of string sizes ensures the generation of realistic and diverse items for analysis.

To generate two sets, \( X_A \) and \( X_B \), each with cardinality \( c \) (the total number of items in each set) and a given Jaccard similarity \( s \) between them, we computed the number of shared items \( \textit{sims} \) and distinct items \( \textit{diffs} \) as follows:

\[
\textit{sims} = \frac{2 \cdot s \cdot c}{1 + s}, \quad \textit{diffs} = c - \textit{sims}
\]

We then constructed the sets by:
\begin{itemize}
    \item Inserting \( \textit{sims} \) shared items in both sets.
    \item Inserting \( \textit{diffs} \) unique items in each set.
\end{itemize}

This ensured that the Jaccard similarity \( s \) between \( X_A \) and \( X_B \) matched the desired value, as given by:

\[
s = \frac{\textit{sims}}{\textit{sims} + 2 \cdot \textit{diffs}}
\]

To maintain readability in the presentation of results, we abbreviated the algorithms that include parameters. The corresponding abbreviations and their full algorithm names are listed in Table \ref{tab:algorithm_abbreviations}.

\begin{table}[h]
\centering
\begin{tabular}{|l|l|}
\hline
\textbf{Algorithm} & \textbf{Abbreviation} \\
\hline
Bloom & \(\mathit{Bl}\) \\
Bucketing & \(\mathit{Bu}\) \\
Rateless & \(\mathit{Ra}\) \\
Bloom + Bucketing & \(\mathit{BlBu}\) \\
Bloom + Rateless & \(\mathit{BlRa}\) \\
Bucketing + Rateless & \(\mathit{BuRa}\) \\
Bloom + Bucketing + Rateless & \(\mathit{BlBuRa}\) \\
\hline
\end{tabular}
\caption{Algorithm Abbreviations}
\label{tab:algorithm_abbreviations}
\end{table}

\subsection{Results}
This subsection presents the results of the evaluation. Note that the y-axis scales in the plots were automatically adjusted based on the data range, so they vary across figures (e.g., from kilobytes to megabytes). This should be kept in mind when comparing plots.

\subsubsection{Approaches without bloom filters}
\begin{figure*}
    \centering
    \includegraphics[width=\textwidth]{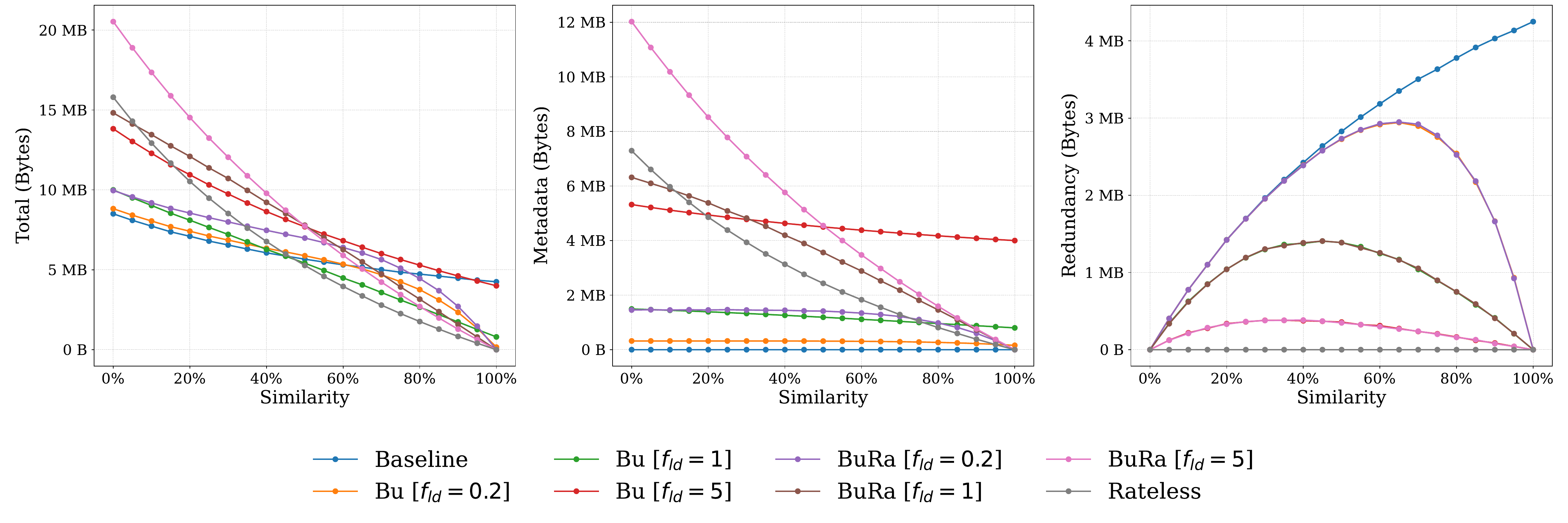}
    \caption{Transmission analysis w.r.t. similarity between a pair of GSets - Bucketing}
    \label{fig:bucketing_transmitted}
\end{figure*}

Figure \ref{fig:bucketing_transmitted} presents transmission analysis for four approaches: \( \mathit{Bu} \), \( \mathit{Ra}\), \( \mathit{BuRa} \), and the Baseline algorithm for comparison. As mentioned in Subsection \ref{section:state-driven-sync}, the state-driven synchronization algorithm represents the baseline.

The redundancy patterns are identical for \( \mathit{Bu} \) and \( \mathit{BuRa} \), with both transmitting the same amount of redundancy for a given similarity. Redundancy starts to increase as similarity increases within each bucket until buckets become identical, at which point redundancy begins to decrease. This behavior is influenced by \( f_{ld} \), with higher values leading to more granular synchronization and less redundancy. The amount of transmitted redundancy and the similarity at which redundancy peaks both decrease with \( f_{ld} \), as higher \( f_{ld} \) results in more granular synchronization and fewer similar-but-not-identical buckets. No redundancy is transmitted by the \( \mathit{Ra} \) algorithm, as only join decompositions with non-matching hashes are exchanged, ensuring all transmitted data is necessary.

The bucketing approaches (i.e.  \( \mathit{Bu} \) and \( \mathit{BuRa} \)),  exchange metadata (hashes or rateless IBLTs) to detect mismatching buckets, followed by sending their indices. In both cases, the number of indices decreases as similarity increases. In the \( \mathit{Bu} \) algorithm, the metadata required to detect mismatching buckets remains constant, while in \( \mathit{BuRa} \), it scales with the number of mismatching buckets, yielding greater metadata reduction as similarity grows. 

The \( \mathit{BuRa} \) approach, however, introduces two sources of overhead: the small size of the \( idSum \) field leads to overhead from other fields in coded symbols, and the symmetric set difference is twice the number of mismatching buckets, causing more metadata to be transmitted when many buckets differ. Consequently, \( \mathit{BuRa} \) results in higher metadata transmission compared to \( \mathit{Bu} \) at low similarity levels but lower transmission at high similarity levels.

The total data consists of metadata, redundant state, and the actual state to be transferred. \( \mathit{Ra} \) has higher metadata overhead at low similarity, but this reduces as similarity increases. Notably, from 45\% similarity, the \( \mathit{Ra}\) algorithm achieves the lowest total data transfer, making it the most efficient in cases with somewhat similar replicas.

\subsubsection{Approaches with Bloom Filters}

Among the algorithms that incorporate Bloom filters, the rateless approach benefits the most. This is expected, as previously established: beyond 45\% similarity, the \( \mathit{Ra} \) algorithm achieves the lowest total data transfer. The initial Bloom filter exchange significantly increases the expected similarity between \(X_A^{com}\) and \(X_B^{com}\), even when small filters are used (e.g., \(\epsilon = 25\%\)).

As detailed in Appendix~\ref{appendix:bloom-bucketing-rateless}, the total data exchanged when combining Bloom filters with both bucketing and rateless reconciliation is comparable to \--- or sometimes exceeds \--- that of using Bloom filters with rateless reconciliation alone. This suggests that the additional complexity introduced by bucketing does not provide a significant advantage in terms of communication efficiency.

\begin{figure*}
    \centering
    \includegraphics[width=0.95\textwidth]{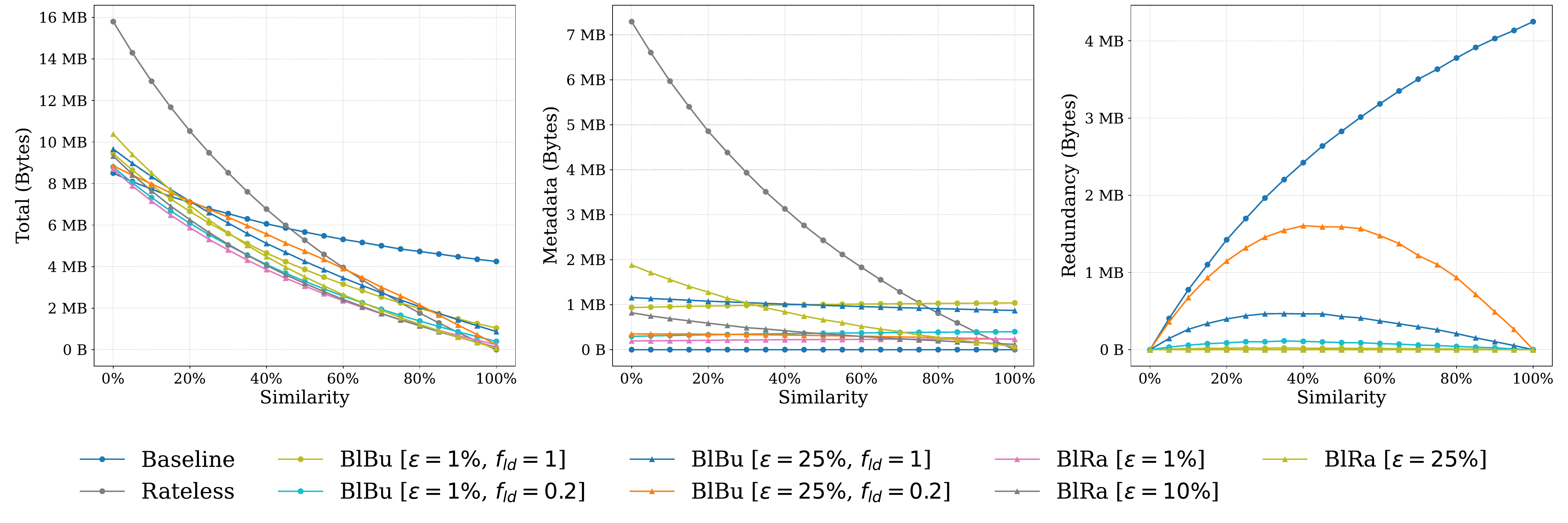}
    \caption{Transmission analysis w.r.t. similarity between a pair of GSets - Bloom}
    \label{fig:bloom_transmitted}
\end{figure*}

All algorithms benefit from incorporating Bloom filters in low-similarity scenarios. As shown in Table~\ref{fig:bloom_transmitted}, the Bloom filter variants of each algorithm consistently outperform their counterparts that do not use Bloom filters. Moreover, the algorithm achieving the lowest total data transfer at any given similarity always includes a Bloom filter, with two notable exceptions. At \(0\%\) similarity, no redundant state is exchanged by any algorithm, and the Baseline approach proves most efficient due to its absence of metadata overhead. At \(100\%\) similarity, the rateless approaches without Bloom filters transmit the least data. This is because the size of the Bloom filter sent from \(A\) to \(B\) is proportional to \(|X_A|\), regardless of similarity. Although Bloom filters typically impose low constant overheads, at very high similarity levels, \(\mathit{Ra}\)’s communication cost—proportional to the symmetric difference—offsets its higher constant factors, making the Bloom-less variant more efficient.

Additionally, Figure~\ref{fig:bloom_transmitted} reveals that at low similarity levels, the best-performing algorithm is the one with the smallest \(\epsilon\). However, as similarity increases, variants with larger \(\epsilon\) values begin to outperform. This behavior stems from the trade-off between Bloom filters and rateless IBLTs: Bloom filter size grows with the set size but benefits from small constants, while rateless IBLTs scale with the symmetric difference and incur larger constants. As a result, Bloom filters with higher \(\epsilon\) values, which require less space, become increasingly advantageous with rising similarity. Eventually, the overhead of Bloom filters can outweigh their benefits, suggesting that omitting them altogether is optimal at high similarity levels. This trend is clearly illustrated in Figure~\ref{fig:high-similarity-transmitted}, in Appendix, which focuses on similarities above 90\%.

These findings point to a promising avenue for future research: dynamically estimating the similarity between replicas to adaptively tune the Bloom filter size, optimizing communication efficiency across a wider range of similarity scenarios.

\subsection{Overall Comparison}

In this subsection, we present a comparison of the best configurations for the most effective algorithms. As demonstrated earlier, incorporating Bloom filters for an initial approximate synchronization before the bucketing stage consistently reduces total data transmission for a given \( f_{ld} \), making this approach a clear improvement.

Additionally, it was shown that adding bucketing to the \( \mathit{BlRa} \) algorithm provides negligible benefits while increasing complexity. This variant consistently outperforms both the \( \mathit{Bu} \) and \( \mathit{BuRa} \) approaches.

Among the \( \mathit{BlBu} \) variants, the configuration \( \mathit{BlBu}[\epsilon = 1\%, f_{ld} = 0.2] \) achieves the lowest total transmission across all similarity levels. In contrast, for the \( \mathit{BlRa} \) method, no single configuration dominates. The optimal Bloom filter size varies depending on the similarity between the replicas' states. However, the \( \mathit{BlRa}[\epsilon = 1\%] \) configuration offers the best trade-off, as it consistently ranks among the algorithms with the least total communication cost across all similarity levels.

For comparison, we also include the Baseline and \(\mathit{Ra}\) approaches. The Baseline represents the current best-performing method for state-based CRDT synchronization that does not rely on external metadata. The \(\mathit{Ra}\) approach represents the simplest adaptation of the state-of-the-art technique for set reconciliation, originally designed for fixed-size elements, applied to variable-sized elements.

\begin{figure*}
    \centering
    \includegraphics[width=\textwidth]{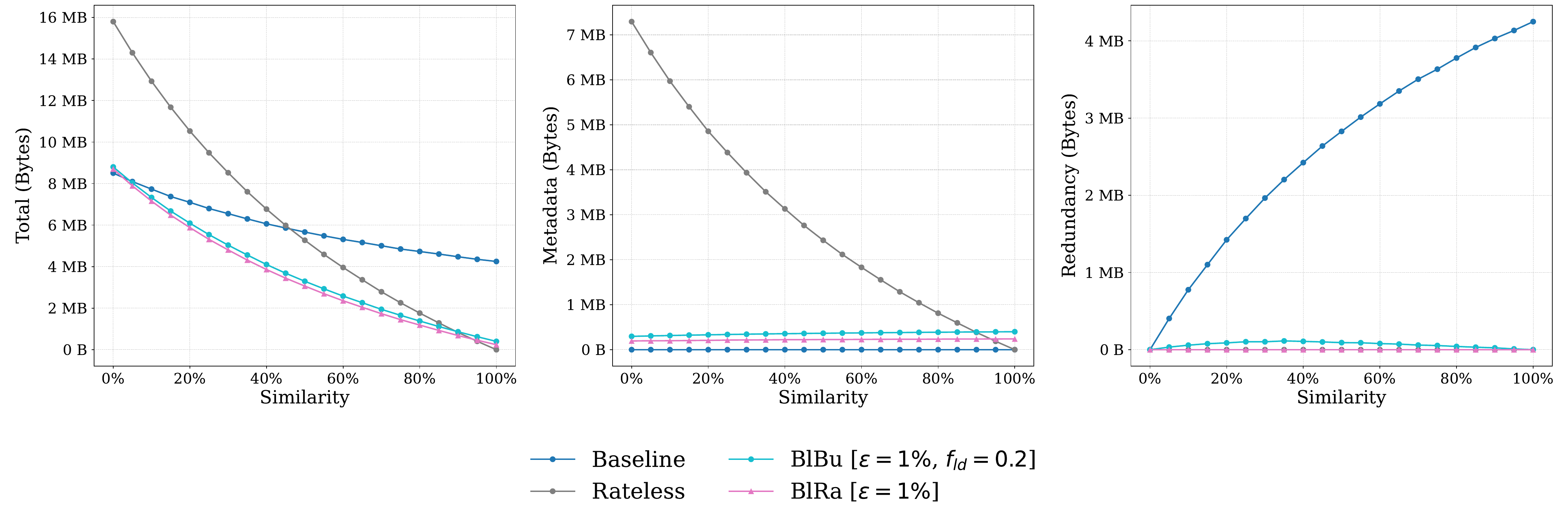}
    \caption{Transmission analysis w.r.t. similarity between a pair of GSets - Best}
    \label{fig:best-transmitted}
\end{figure*}

\begin{figure*}
    \centering
    \includegraphics[width=\textwidth]{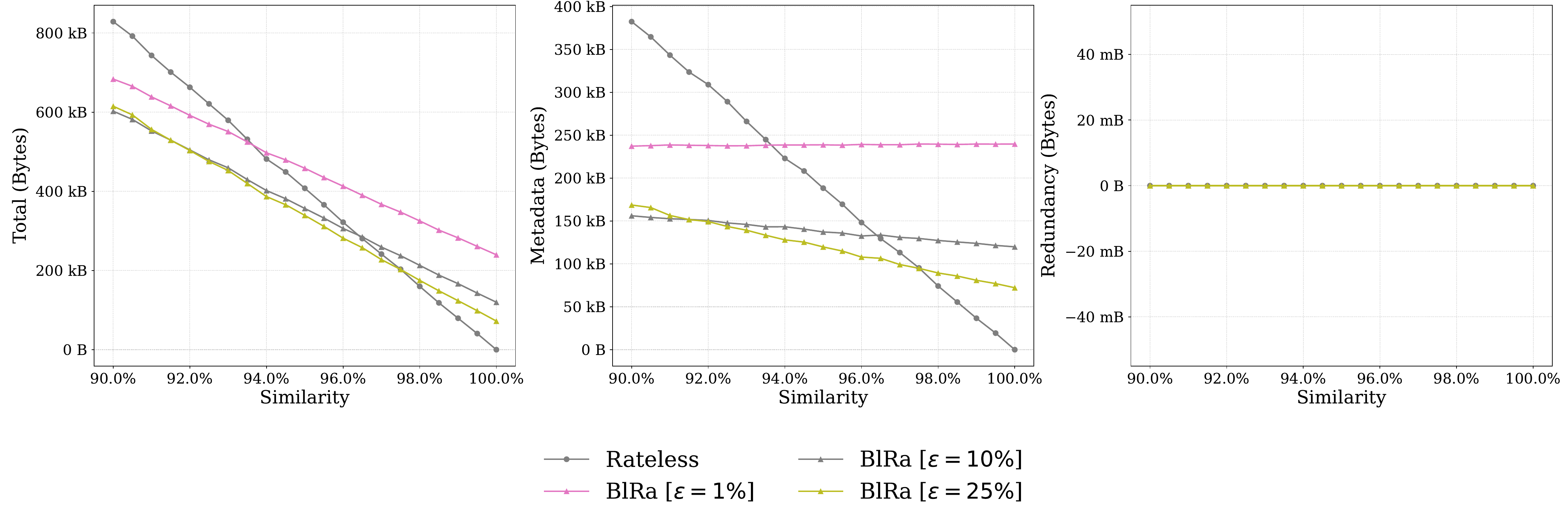}
    \caption{Transmission analysis w.r.t. similarity between a pair of GSets - High Similarity}
    \label{fig:high-similarity-transmitted}
\end{figure*}

Figure~\ref{fig:best-transmitted} compares these 4 methods. It is clear that our novel algorithms outperform the Baseline. Additionally, for similarities below \(50\%\), the \(\mathit{Ra}\) approach performs worse than the Baseline, and only at higher similarity levels does \(\mathit{Ra}\) become competitive with the remaining two methods. The introduction of the bloom filter step is clearly beneficial when the similarity between the two replicas is not close to \(100\%\), as already shown when analyzing analyzing Figure~\ref{fig:high-similarity-transmitted}.

Between \( \mathit{BlBu}[\epsilon = 1\%, f_{ld} = 0.2] \) and \( \mathit{BlRa}[\epsilon = 1\%] \), the latter consistently achieves better performance across all similarity levels, exhibiting lower metadata overhead and redundant data transmission. In addition, \( \mathit{BlBu} \) introduces added implementation complexity due to the bucketing step, while \( \mathit{BlRa} \) builds only on well-established Bloom filters and rateless IBLTs. Aside from a minor adaptation to support variable-sized elements, it remains significantly simpler to implement for those familiar with these techniques.

\section{Related Work}
In this work, we assume no prior knowledge from previous synchronization rounds. In contrast, $\delta$-based CRDTs~\cite{DBLP:conf/netys/AlmeidaSB15, DBLP:conf/icde/EnesAB019} synchronize using small incremental states (deltas) and track acknowledgments to avoid resending them, leveraging prior synchronization history. Our approach, by contrast, assumes no knowledge prior to synchronization.

Unlike $\delta$-CRDTs, which require metadata proportional to the number of operations, our method's metadata scales with the number of decompositions (i.e., the state size). While $\delta$-CRDTs may use garbage collection to reduce metadata overhead, premature deletion of deltas can lead to costly full-state transfers, making them less suitable for highly dynamic networks with frequent churn. Nonetheless, they remain effective in low-churn settings and can be integrated with our approach to avoid full-state transmission when deltas are prematurely garbage-collected, enabling more aggressive garbage collection strategies.

$\Delta$-CRDTs~\cite{DBLP:conf/eurosys/LindeLP16} are conceptually closer to our approach, as the initiator first sends metadata—its causal context (typically, a vector clock)—which the receiver uses to compute $\Delta$, the state that the initiator is missing. However, this requires a data-type-specific definition of the $\mathsf{getDelta()}$ function, which computes the minimal state $\Delta$ that the receiver should send to the initiator based on the initiator’s causal context. As a result, state-based CRDT designs need to be modified to support $\Delta$-CRDT synchronization.

In contrast, our synchronization algorithm integrates seamlessly with any state-based CRDT that defines the irredundant join decomposition operator $\Downarrow$, enabling a generic, data-type-agnostic implementation that can be readily applied to existing state-based CRDT designs. Even still, $\Delta$-CRDTs are applicable to highly dynamic communication patterns, require only 3 messages for the synchronization of both replicas, have no redundancy, and the metadata overhead of two vector clocks is minimal. $\Delta$-CRDTs require maintaining additional metadata which can grow indefinitely (e.g., tombstones), which, as in $\delta$-CRDTs, can be garbage collected at the risk of triggering a full state transmission. As with $\delta$-CRDTs, they can be integrated with our approach to avoid full-state transmission and enable more aggressive garbage collection strategies.

Merkle Trees~\cite{merkle_trees} have also been employed in anti-entropy protocols to efficiently identify set differences~\cite{dynamo}. By organizing elements into a hierarchical tree structure, they enable entire subtrees to be skipped when their hashes match. However, this approach requires one message per tree level, resulting in a number of exchanged messages proportional to the logarithm of the set size, even in cases where the sets differ by only a single element.

Graphene~\cite{graphene} combines Bloom filters and IBLTs to efficiently synchronize blocks and transactions in blockchain networks. Since Graphene predates rateless set reconciliation, it relies on regular IBLTs, which provide only probabilistic guarantees: a fixed number of coded symbols must be chosen in advance, allowing decoding up to a certain number of differences with a given probability. In contrast, rateless set reconciliation streams coded symbols until synchronization is guaranteed, with the probabilistic aspect influencing only the number of symbols needed, not the convergence itself.

Graphene was specifically designed for the dissemination of new blocks and transactions in blockchain systems. Given the delay between a transaction's broadcast and its inclusion in a mined block, most nodes already possess the majority of transactions included in a new block. To exploit this, Graphene first reconciles transaction identifiers (cryptographic hashes) and then requests any missing transactions. This strategy, already present in prior blockchain propagation literature, optimizes synchronization by avoiding the transmission of full transactions. However, Graphene does not address the challenge of variable-sized elements in set reconciliation. In contrast, our approach generalizes set reconciliation to handle elements of varying sizes, which is crucial for applying it to state-based CRDTs.

Additionally, Graphene provides a comprehensive theoretical analysis of how to jointly parameterize the Bloom filter and the IBLT to minimize their total size while maintaining a high, tunable success rate. The success probability can be set arbitrarily high at the cost of overprovisioning the Bloom filter and IBLT. This analysis could be leveraged in future extensions of our approach, particularly to minimize the summed size of the Bloom filter and the expected size of the rateless IBLT.

\section{Conclusion and Future Work}

We designed, implemented, and evaluated \textit{ConflictSync}, a digest-driven synchronization protocol for state-based CRDTs that leverages irredundant join decompositions. ConflictSync reframes the problem of CRDT synchronization as one of set reconciliation, enabling replicas to exchange compact digests and reconcile differences with near-minimal communication overhead.

A key strength of ConflictSync lies in its versatility: it applies to any state-based CRDT, requires no prior synchronization history or external metadata, and achieves low transmission overhead across a wide range of similarity levels. Our experiments, conducted on GSet CRDTs in controlled environments, show that ConflictSync significantly reduces total data transfer—achieving up to an 18$\times$ improvement over traditional state-based synchronization in high-similarity scenarios. While our evaluation focused on GSets, further work is needed to validate ConflictSync with more complex data types, such as nested maps.

We highlight several promising directions for advancing \textit{ConflictSync} in the context of pairwise synchronization: (i) dynamically tuning Bloom filter sizes based on observed or estimated similarity, and (ii) exploring hierarchical digest structures such as Merkle trees. 

In addition, our results suggest the potential to extend \textit{ConflictSync} to multiparty synchronization. A straightforward approach could involve periodically selecting a random neighbor for synchronization, providing a useful benchmark against current anti-entropy methods for CRDTs. To further optimize this process, synchronizing with multiple peers concurrently could reduce latency. Additionally, incorporating topology awareness \--- where the synchronization strategy adapts to the network structure \--- or factoring in state similarity when selecting synchronization partners could significantly enhance efficiency, minimizing unnecessary data transfer and improving overall synchronization performance.

\bibliography{main}

\section{Appendix}


\subsection{Evaluation}

\begin{table*}[h]
	\centering
	\begin{tabular}{lccccccc}
		\toprule
		\textbf{Algorithm} & \textbf{0\%} & \textbf{25\%} & \textbf{50\%} & \textbf{75\%} & \textbf{90\%} & \textbf{95\%} & \textbf{100\%} \\
		\midrule
		Baseline & \textbf{\SI{8.5}{\mega\byte}} & \SI{6.8}{\mega\byte} & \SI{5.67}{\mega\byte} & \SI{4.85}{\mega\byte} & \SI{4.48}{\mega\byte} & \SI{4.35}{\mega\byte} & \SI{4.25}{\mega\byte} \\
		Bu [$f_{ld} = 0.2$] & \SI{8.82}{\mega\byte} & \SI{7.11}{\mega\byte} & \SI{5.88}{\mega\byte} & \SI{4.25}{\mega\byte} & \SI{2.34}{\mega\byte} & \SI{1.35}{\mega\byte} & \SI{160}{\kilo\byte} \\
		Bu [$f_{ld} = 1$] & \SI{10}{\mega\byte} & \SI{7.65}{\mega\byte} & \SI{5.41}{\mega\byte} & \SI{3.11}{\mega\byte} & \SI{1.74}{\mega\byte} & \SI{1.26}{\mega\byte} & \SI{800}{\kilo\byte} \\
		Bu [$f_{ld} = 5$] & \SI{13.82}{\mega\byte} & \SI{10.31}{\mega\byte} & \SI{7.69}{\mega\byte} & \SI{5.64}{\mega\byte} & \SI{4.61}{\mega\byte} & \SI{4.3}{\mega\byte} & \SI{4}{\mega\byte} \\
		BuRa [$f_{ld} = 0.2$] & \SI{9.96}{\mega\byte} & \SI{8.26}{\mega\byte} & \SI{6.99}{\mega\byte} & \SI{5.09}{\mega\byte} & \SI{2.71}{\mega\byte} & \SI{1.47}{\mega\byte} & \textbf{\SI{24}{\byte}} \\
		BuRa [$f_{ld} = 1$] & \SI{14.82}{\mega\byte} & \SI{11.38}{\mega\byte} & \SI{7.79}{\mega\byte} & \SI{3.92}{\mega\byte} & \SI{1.58}{\mega\byte} & \SI{792.6}{\kilo\byte} & \textbf{\SI{24}{\byte}} \\
		BuRa [$f_{ld} = 5$] & \SI{20.53}{\mega\byte} & \SI{13.24}{\mega\byte} & \SI{7.73}{\mega\byte} & \SI{3.45}{\mega\byte} & \SI{1.28}{\mega\byte} & \SI{633.6}{\kilo\byte} & \textbf{\SI{24}{\byte}} \\
		Rateless & \SI{15.8}{\mega\byte} & \SI{9.48}{\mega\byte} & \SI{5.27}{\mega\byte} & \SI{2.26}{\mega\byte} & \SI{829.6}{\kilo\byte} & \SI{405.1}{\kilo\byte} & \textbf{\SI{24}{\byte}} \\
		BlBu [$\epsilon = 1\%$, $f_{ld} = 1$] & \SI{9.44}{\mega\byte} & \SI{6.1}{\mega\byte} & \SI{3.86}{\mega\byte} & \SI{2.25}{\mega\byte} & \SI{1.48}{\mega\byte} & \SI{1.26}{\mega\byte} & \SI{1.04}{\mega\byte} \\
		BlBu [$\epsilon = 1\%$, $f_{ld} = 0.2$] & \SI{8.8}{\mega\byte} & \SI{5.54}{\mega\byte} & \SI{3.29}{\mega\byte} & \SI{1.65}{\mega\byte} & \SI{863.5}{\kilo\byte} & \SI{625.2}{\kilo\byte} & \SI{399.7}{\kilo\byte} \\
		BlBu [$\epsilon = 25\%$, $f_{ld} = 1$] & \SI{9.66}{\mega\byte} & \SI{6.6}{\mega\byte} & \SI{4.25}{\mega\byte} & \SI{2.4}{\mega\byte} & \SI{1.44}{\mega\byte} & \SI{1.15}{\mega\byte} & \SI{872.2}{\kilo\byte} \\
		BlBu [$\epsilon = 25\%$, $f_{ld} = 0.2$] & \SI{8.86}{\mega\byte} & \SI{6.76}{\mega\byte} & \SI{4.74}{\mega\byte} & \SI{2.59}{\mega\byte} & \SI{1.19}{\mega\byte} & \SI{725.1}{\kilo\byte} & \SI{232.2}{\kilo\byte} \\
		BlBuRa [$\epsilon = 1\%$, $f_{ld} = 1$] & \SI{8.77}{\mega\byte} & \SI{5.38}{\mega\byte} & \SI{3.11}{\mega\byte} & \SI{1.47}{\mega\byte} & \SI{691}{\kilo\byte} & \SI{462.2}{\kilo\byte} & \SI{239.7}{\kilo\byte} \\
		BlBuRa [$\epsilon = 1\%$, $f_{ld} = 0.2$] & \SI{8.77}{\mega\byte} & \SI{5.46}{\mega\byte} & \SI{3.18}{\mega\byte} & \SI{1.51}{\mega\byte} & \SI{712.4}{\kilo\byte} & \SI{468.3}{\kilo\byte} & \SI{239.7}{\kilo\byte} \\
		BlBuRa [$\epsilon = 25\%$, $f_{ld} = 1$] & \SI{11.41}{\mega\byte} & \SI{7.47}{\mega\byte} & \SI{4.43}{\mega\byte} & \SI{2.02}{\mega\byte} & \SI{819.5}{\kilo\byte} & \SI{431.8}{\kilo\byte} & \SI{72.24}{\kilo\byte} \\
		BlBuRa [$\epsilon = 25\%$, $f_{ld} = 0.2$] & \SI{9.88}{\mega\byte} & \SI{7.59}{\mega\byte} & \SI{5.33}{\mega\byte} & \SI{2.8}{\mega\byte} & \SI{1.19}{\mega\byte} & \SI{636.9}{\kilo\byte} & \SI{72.24}{\kilo\byte} \\
		BlRa [$\epsilon = 1\%$] & \SI{8.7}{\mega\byte} & \textbf{\SI{5.31}{\mega\byte}} & \textbf{\SI{3.06}{\mega\byte}} & \SI{1.45}{\mega\byte} & \SI{681.9}{\kilo\byte} & \SI{456}{\kilo\byte} & \SI{239.7}{\kilo\byte} \\
		BlRa [$\epsilon = 10\%$] & \SI{9.33}{\mega\byte} & \SI{5.64}{\mega\byte} & \SI{3.19}{\mega\byte} & \textbf{\SI{1.43}{\mega\byte}} & \textbf{\SI{601.4}{\kilo\byte}} & \SI{354.7}{\kilo\byte} & \SI{119.9}{\kilo\byte} \\
		BlRa [$\epsilon = 25\%$] & \SI{10.39}{\mega\byte} & \SI{6.24}{\mega\byte} & \SI{3.5}{\mega\byte} & \SI{1.54}{\mega\byte} & \SI{612.1}{\kilo\byte} & \textbf{\SI{337.4}{\kilo\byte}} & \SI{72.24}{\kilo\byte} \\
		\bottomrule
	\end{tabular}
	\caption{GSet - Transmitted Total}
	\label{tab:gset_transmitted_total}
\end{table*}

\begin{table*}[h]
	\centering
	\begin{tabular}{lccccccc}
		\toprule
		\textbf{Algorithm} & \textbf{0\%} & \textbf{25\%} & \textbf{50\%} & \textbf{75\%} & \textbf{90\%} & \textbf{95\%} & \textbf{100\%} \\
		\midrule
		Baseline & \SI{0}{\byte} & \SI{0}{\byte} & \SI{0}{\byte} & \SI{0}{\byte} & \SI{0}{\byte} & \SI{0}{\byte} & \SI{0}{\byte} \\
		Bu [$f_{ld} = 0.2$] & \SI{320}{\kilo\byte} & \SI{319.6}{\kilo\byte} & \SI{314.2}{\kilo\byte} & \SI{281.3}{\kilo\byte} & \SI{226}{\kilo\byte} & \SI{195.9}{\kilo\byte} & \SI{160}{\kilo\byte} \\
		Bu [$f_{ld} = 1$] & \SI{1.49}{\mega\byte} & \SI{1.36}{\mega\byte} & \SI{1.19}{\mega\byte} & \SI{998.1}{\kilo\byte} & \SI{879.9}{\kilo\byte} & \SI{840}{\kilo\byte} & \SI{800}{\kilo\byte} \\
		Bu [$f_{ld} = 5$] & \SI{5.32}{\mega\byte} & \SI{4.85}{\mega\byte} & \SI{4.5}{\mega\byte} & \SI{4.22}{\mega\byte} & \SI{4.08}{\mega\byte} & \SI{4.04}{\mega\byte} & \SI{4}{\mega\byte} \\
		BuRa [$f_{ld} = 0.2$] & \SI{1.46}{\mega\byte} & \SI{1.47}{\mega\byte} & \SI{1.41}{\mega\byte} & \SI{1.1}{\mega\byte} & \SI{600.2}{\kilo\byte} & \SI{331.2}{\kilo\byte} & \SI{24}{\byte} \\
		BuRa [$f_{ld} = 1$] & \SI{6.32}{\mega\byte} & \SI{5.09}{\mega\byte} & \SI{3.56}{\mega\byte} & \SI{1.81}{\mega\byte} & \SI{728.2}{\kilo\byte} & \SI{366.3}{\kilo\byte} & \SI{24}{\byte} \\
		BuRa [$f_{ld} = 5$] & \SI{12.03}{\mega\byte} & \SI{7.78}{\mega\byte} & \SI{4.55}{\mega\byte} & \SI{2.03}{\mega\byte} & \SI{760.2}{\kilo\byte} & \SI{373.9}{\kilo\byte} & \SI{24}{\byte} \\
		Rateless & \SI{7.29}{\mega\byte} & \SI{4.38}{\mega\byte} & \SI{2.43}{\mega\byte} & \SI{1.04}{\mega\byte} & \SI{384.6}{\kilo\byte} & \SI{187.2}{\kilo\byte} & \SI{24}{\byte} \\
		BlBu [$\epsilon = 1\%$, $f_{ld} = 1$] & \SI{937.7}{\kilo\byte} & \SI{978}{\kilo\byte} & \SI{1.01}{\mega\byte} & \SI{1.03}{\mega\byte} & \SI{1.03}{\mega\byte} & \SI{1.04}{\mega\byte} & \SI{1.04}{\mega\byte} \\
		BlBu [$\epsilon = 1\%$, $f_{ld} = 0.2$] & \SI{296.2}{\kilo\byte} & \SI{338.1}{\kilo\byte} & \SI{365.5}{\kilo\byte} & \SI{385.2}{\kilo\byte} & \SI{394.4}{\kilo\byte} & \SI{397.1}{\kilo\byte} & \SI{399.7}{\kilo\byte} \\
		BlBu [$\epsilon = 25\%$, $f_{ld} = 1$] & \SI{1.16}{\mega\byte} & \SI{1.06}{\mega\byte} & \SI{985.1}{\kilo\byte} & \SI{924.4}{\kilo\byte} & \SI{891.1}{\kilo\byte} & \SI{881.9}{\kilo\byte} & \SI{872.2}{\kilo\byte} \\
		BlBu [$\epsilon = 25\%$, $f_{ld} = 0.2$] & \SI{351.5}{\kilo\byte} & \SI{340.4}{\kilo\byte} & \SI{313.4}{\kilo\byte} & \SI{276.7}{\kilo\byte} & \SI{250.4}{\kilo\byte} & \SI{241.9}{\kilo\byte} & \SI{232.2}{\kilo\byte} \\
		BlBuRa [$\epsilon = 1\%$, $f_{ld} = 1$] & \SI{267.3}{\kilo\byte} & \SI{255.9}{\kilo\byte} & \SI{254.8}{\kilo\byte} & \SI{244.3}{\kilo\byte} & \SI{241.1}{\kilo\byte} & \SI{241.3}{\kilo\byte} & \SI{239.7}{\kilo\byte} \\
		BlBuRa [$\epsilon = 1\%$, $f_{ld} = 0.2$] & \SI{267.3}{\kilo\byte} & \SI{257}{\kilo\byte} & \SI{248.9}{\kilo\byte} & \SI{243.1}{\kilo\byte} & \SI{242.5}{\kilo\byte} & \SI{240.3}{\kilo\byte} & \SI{239.7}{\kilo\byte} \\
		BlBuRa [$\epsilon = 25\%$, $f_{ld} = 1$] & \SI{2.9}{\mega\byte} & \SI{1.94}{\mega\byte} & \SI{1.17}{\mega\byte} & \SI{561.4}{\kilo\byte} & \SI{265.6}{\kilo\byte} & \SI{160.4}{\kilo\byte} & \SI{72.24}{\kilo\byte} \\
		BlBuRa [$\epsilon = 25\%$, $f_{ld} = 0.2$] & \SI{1.38}{\mega\byte} & \SI{1.18}{\mega\byte} & \SI{892.9}{\kilo\byte} & \SI{503.2}{\kilo\byte} & \SI{251.1}{\kilo\byte} & \SI{162}{\kilo\byte} & \SI{72.24}{\kilo\byte} \\
		BlRa [$\epsilon = 1\%$] & \SI{195.3}{\kilo\byte} & \SI{213.7}{\kilo\byte} & \SI{226.4}{\kilo\byte} & \SI{232.4}{\kilo\byte} & \SI{236.9}{\kilo\byte} & \SI{238.1}{\kilo\byte} & \SI{239.7}{\kilo\byte} \\
		BlRa [$\epsilon = 10\%$] & \SI{822.3}{\kilo\byte} & \SI{541.1}{\kilo\byte} & \SI{353.1}{\kilo\byte} & \SI{220}{\kilo\byte} & \SI{156.4}{\kilo\byte} & \SI{136.8}{\kilo\byte} & \SI{119.9}{\kilo\byte} \\
		BlRa [$\epsilon = 25\%$] & \SI{1.88}{\mega\byte} & \SI{1.14}{\mega\byte} & \SI{666.9}{\kilo\byte} & \SI{328}{\kilo\byte} & \SI{167.1}{\kilo\byte} & \SI{119.5}{\kilo\byte} & \SI{72.24}{\kilo\byte} \\
		\bottomrule
	\end{tabular}
	\caption{GSet - Transmitted Metadata}
	\label{tab:gset_transmitted_metadata}
\end{table*}

\begin{table*}[h]
	\centering
	\begin{tabular}{lccccccc}
		\toprule
		\textbf{Algorithm} & \textbf{0\%} & \textbf{25\%} & \textbf{50\%} & \textbf{75\%} & \textbf{90\%} & \textbf{95\%} & \textbf{100\%} \\
		\midrule
		Baseline & \SI{0}{\byte} & \SI{1.7}{\mega\byte} & \SI{2.83}{\mega\byte} & \SI{3.63}{\mega\byte} & \SI{4.03}{\mega\byte} & \SI{4.13}{\mega\byte} & \SI{4.25}{\mega\byte} \\
		Bu [$f_{ld} = 0.2$] & \SI{0}{\byte} & \SI{1.7}{\mega\byte} & \SI{2.73}{\mega\byte} & \SI{2.75}{\mega\byte} & \SI{1.66}{\mega\byte} & \SI{932.3}{\kilo\byte} & \SI{0}{\byte} \\
		Bu [$f_{ld} = 1$] & \SI{0}{\byte} & \SI{1.19}{\mega\byte} & \SI{1.39}{\mega\byte} & \SI{895.2}{\kilo\byte} & \SI{412.1}{\kilo\byte} & \SI{206.5}{\kilo\byte} & \SI{0}{\byte} \\
		Bu [$f_{ld} = 5$] & \SI{0}{\byte} & \SI{361.2}{\kilo\byte} & \SI{357.3}{\kilo\byte} & \SI{205.2}{\kilo\byte} & \SI{85.89}{\kilo\byte} & \SI{42.1}{\kilo\byte} & \SI{0}{\byte} \\
		BuRa [$f_{ld} = 0.2$] & \SI{0}{\byte} & \SI{1.7}{\mega\byte} & \SI{2.73}{\mega\byte} & \SI{2.78}{\mega\byte} & \SI{1.66}{\mega\byte} & \SI{923.5}{\kilo\byte} & \SI{0}{\byte} \\
		BuRa [$f_{ld} = 1$] & \SI{0}{\byte} & \SI{1.19}{\mega\byte} & \SI{1.39}{\mega\byte} & \SI{898.5}{\kilo\byte} & \SI{406.9}{\kilo\byte} & \SI{208.4}{\kilo\byte} & \SI{0}{\byte} \\
		BuRa [$f_{ld} = 5$] & \SI{0}{\byte} & \SI{363.8}{\kilo\byte} & \SI{347.8}{\kilo\byte} & \SI{200.6}{\kilo\byte} & \SI{78.61}{\kilo\byte} & \SI{41.8}{\kilo\byte} & \SI{0}{\byte} \\
		Rateless & \SI{0}{\byte} & \SI{0}{\byte} & \SI{0}{\byte} & \SI{0}{\byte} & \SI{0}{\byte} & \SI{0}{\byte} & \SI{0}{\byte} \\
		BlBu [$\epsilon = 1\%$, $f_{ld} = 1$] & \SI{0}{\byte} & \SI{22.25}{\kilo\byte} & \SI{19.36}{\kilo\byte} & \SI{11.04}{\kilo\byte} & \SI{4.84}{\kilo\byte} & \SI{2.6}{\kilo\byte} & \SI{0}{\byte} \\
		BlBu [$\epsilon = 1\%$, $f_{ld} = 0.2$] & \SI{0}{\byte} & \SI{102.9}{\kilo\byte} & \SI{91.58}{\kilo\byte} & \SI{54.04}{\kilo\byte} & \SI{24.08}{\kilo\byte} & \SI{10.22}{\kilo\byte} & \SI{0}{\byte} \\
		BlBu [$\epsilon = 25\%$, $f_{ld} = 1$] & \SI{0}{\byte} & \SI{440.3}{\kilo\byte} & \SI{430.1}{\kilo\byte} & \SI{258.2}{\kilo\byte} & \SI{104.1}{\kilo\byte} & \SI{53.95}{\kilo\byte} & \SI{0}{\byte} \\
		BlBu [$\epsilon = 25\%$, $f_{ld} = 0.2$] & \SI{0}{\byte} & \SI{1.32}{\mega\byte} & \SI{1.59}{\mega\byte} & \SI{1.1}{\mega\byte} & \SI{490.7}{\kilo\byte} & \SI{265.3}{\kilo\byte} & \SI{0}{\byte} \\
		BlBuRa [$\epsilon = 1\%$, $f_{ld} = 1$] & \SI{0}{\byte} & \SI{20.82}{\kilo\byte} & \SI{21.55}{\kilo\byte} & \SI{10}{\kilo\byte} & \SI{4.9}{\kilo\byte} & \SI{2.95}{\kilo\byte} & \SI{0}{\byte} \\
		BlBuRa [$\epsilon = 1\%$, $f_{ld} = 0.2$] & \SI{0}{\byte} & \SI{102.3}{\kilo\byte} & \SI{93.84}{\kilo\byte} & \SI{52.71}{\kilo\byte} & \SI{24.94}{\kilo\byte} & \SI{10.13}{\kilo\byte} & \SI{0}{\byte} \\
		BlBuRa [$\epsilon = 25\%$, $f_{ld} = 1$] & \SI{0}{\byte} & \SI{433.7}{\kilo\byte} & \SI{420.5}{\kilo\byte} & \SI{245.7}{\kilo\byte} & \SI{108.8}{\kilo\byte} & \SI{53.48}{\kilo\byte} & \SI{0}{\byte} \\
		BlBuRa [$\epsilon = 25\%$, $f_{ld} = 0.2$] & \SI{0}{\byte} & \SI{1.31}{\mega\byte} & \SI{1.6}{\mega\byte} & \SI{1.08}{\mega\byte} & \SI{498.3}{\kilo\byte} & \SI{257}{\kilo\byte} & \SI{0}{\byte} \\
		BlRa [$\epsilon = 1\%$] & \SI{0}{\byte} & \SI{0}{\byte} & \SI{0}{\byte} & \SI{0}{\byte} & \SI{0}{\byte} & \SI{0}{\byte} & \SI{0}{\byte} \\
		BlRa [$\epsilon = 10\%$] & \SI{0}{\byte} & \SI{0}{\byte} & \SI{0}{\byte} & \SI{0}{\byte} & \SI{0}{\byte} & \SI{0}{\byte} & \SI{0}{\byte} \\
		BlRa [$\epsilon = 25\%$] & \SI{0}{\byte} & \SI{0}{\byte} & \SI{0}{\byte} & \SI{0}{\byte} & \SI{0}{\byte} & \SI{0}{\byte} & \SI{0}{\byte} \\
		\bottomrule
	\end{tabular}
	\caption{GSet - Transmitted Redundancy}
	\label{tab:gset_transmitted_redundancy}
\end{table*}

\begin{table*}[h]
	\centering
	\begin{tabular}{lccccccc}
		\toprule
		\textbf{Algorithm} & \textbf{0\%} &\textbf{25\%} &\textbf{50\%} &\textbf{75\%} &\textbf{90\%} &\textbf{95\%} &\textbf{100\%} \\
		\midrule
		Baseline & 0.0\% &0.0\% &0.0\% &0.0\% &0.0\% &0.0\% &0.0\% \\
		Bu [$f_{ld} = 0.2$] & 3.6\% &4.5\% &5.3\% &6.6\% &9.7\% &14.6\% &100.0\% \\
		Bu [$f_{ld} = 1$] & 14.9\% &17.8\% &22.0\% &32.1\% &50.7\% &66.4\% &100.0\% \\
		Bu [$f_{ld} = 5$] & 38.5\% &47.1\% &58.5\% &74.8\% &88.5\% &94.0\% &100.0\% \\
		BuRa [$f_{ld} = 0.2$] & 14.6\% &17.7\% &20.2\% &21.7\% &22.2\% &22.5\% &100.0\% \\
		BuRa [$f_{ld} = 1$] & 42.6\% &44.7\% &45.7\% &46.1\% &46.1\% &46.2\% &100.0\% \\
		BuRa [$f_{ld} = 5$] & 58.6\% &58.8\% &58.8\% &58.9\% &59.2\% &59.0\% &100.0\% \\
		Rateless & 46.2\% &46.2\% &46.2\% &46.2\% &46.4\% &46.2\% &100.0\% \\
		BlBu [$\epsilon = 1\%$, $f_{ld} = 1$] & 9.9\% &16.0\% &26.0\% &45.5\% &69.7\% &82.5\% &100.0\% \\
		BlBu [$\epsilon = 1\%$, $f_{ld} = 0.2$] & 3.4\% &6.1\% &11.1\% &23.3\% &45.7\% &63.5\% &100.0\% \\
		BlBu [$\epsilon = 25\%$, $f_{ld} = 1$] & 12.0\% &16.1\% &23.2\% &38.6\% &61.9\% &76.4\% &100.0\% \\
		BlBu [$\epsilon = 25\%$, $f_{ld} = 0.2$] & 4.0\% &5.0\% &6.6\% &10.7\% &21.1\% &33.4\% &100.0\% \\
		BlBuRa [$\epsilon = 1\%$, $f_{ld} = 1$] & 3.0\% &4.8\% &8.2\% &16.6\% &34.9\% &52.2\% &100.0\% \\
		BlBuRa [$\epsilon = 1\%$, $f_{ld} = 0.2$] & 3.0\% &4.7\% &7.8\% &16.1\% &34.0\% &51.3\% &100.0\% \\
		BlBuRa [$\epsilon = 25\%$, $f_{ld} = 1$] & 25.4\% &26.0\% &26.5\% &27.8\% &32.4\% &37.1\% &100.0\% \\
		BlBuRa [$\epsilon = 25\%$, $f_{ld} = 0.2$] & 14.0\% &15.6\% &16.7\% &18.0\% &21.0\% &25.4\% &100.0\% \\
		BlRa [$\epsilon = 1\%$] & 2.2\% &4.0\% &7.4\% &16.1\% &34.7\% &52.2\% &100.0\% \\
		BlRa [$\epsilon = 10\%$] & 8.8\% &9.6\% &11.1\% &15.3\% &26.0\% &38.6\% &100.0\% \\
		BlRa [$\epsilon = 25\%$] & 18.1\% &18.3\% &19.0\% &21.3\% &27.3\% &35.4\% &100.0\% \\
		\bottomrule
	\end{tabular}
	\caption{GSet - Metadata Ratios}
	\label{tab:gset_metadata_ratios}
\end{table*}

\begin{table*}[h]
	\centering
	\begin{tabular}{lccccccc}
		\toprule
		\textbf{Algorithm} & \textbf{0\%} &\textbf{25\%} &\textbf{50\%} &\textbf{75\%} &\textbf{90\%} &\textbf{95\%} &\textbf{100\%} \\
		\midrule
		Baseline & 0.0\% &25.0\% &49.9\% &74.9\% &90.1\% &95.0\% &100.0\% \\
		Bu [$f_{ld} = 0.2$] & 0.0\% &23.8\% &46.4\% &64.8\% &71.3\% &69.3\% &0.0\% \\
		Bu [$f_{ld} = 1$] & 0.0\% &15.6\% &25.6\% &28.8\% &23.7\% &16.3\% &0.0\% \\
		Bu [$f_{ld} = 5$] & 0.0\% &3.5\% &4.6\% &3.6\% &1.9\% &1.0\% &0.0\% \\
		BuRa [$f_{ld} = 0.2$] & 0.0\% &20.5\% &39.1\% &54.5\% &61.4\% &62.7\% &0.0\% \\
		BuRa [$f_{ld} = 1$] & 0.0\% &10.5\% &17.8\% &22.9\% &25.8\% &26.3\% &0.0\% \\
		BuRa [$f_{ld} = 5$] & 0.0\% &2.7\% &4.5\% &5.8\% &6.1\% &6.6\% &0.0\% \\
		Rateless & 0.0\% &0.0\% &0.0\% &0.0\% &0.0\% &0.0\% &0.0\% \\
		BlBu [$\epsilon = 1\%$, $f_{ld} = 1$] & 0.0\% &0.4\% &0.5\% &0.5\% &0.3\% &0.2\% &0.0\% \\
		BlBu [$\epsilon = 1\%$, $f_{ld} = 0.2$] & 0.0\% &1.9\% &2.8\% &3.3\% &2.8\% &1.6\% &0.0\% \\
		BlBu [$\epsilon = 25\%$, $f_{ld} = 1$] & 0.0\% &6.7\% &10.1\% &10.8\% &7.2\% &4.7\% &0.0\% \\
		BlBu [$\epsilon = 25\%$, $f_{ld} = 0.2$] & 0.0\% &19.5\% &33.5\% &42.5\% &41.4\% &36.6\% &0.0\% \\
		BlBuRa [$\epsilon = 1\%$, $f_{ld} = 1$] & 0.0\% &0.4\% &0.7\% &0.7\% &0.7\% &0.6\% &0.0\% \\
		BlBuRa [$\epsilon = 1\%$, $f_{ld} = 0.2$] & 0.0\% &1.9\% &3.0\% &3.5\% &3.5\% &2.2\% &0.0\% \\
		BlBuRa [$\epsilon = 25\%$, $f_{ld} = 1$] & 0.0\% &5.8\% &9.5\% &12.2\% &13.3\% &12.4\% &0.0\% \\
		BlBuRa [$\epsilon = 25\%$, $f_{ld} = 0.2$] & 0.0\% &17.3\% &30.1\% &38.6\% &41.7\% &40.3\% &0.0\% \\
		BlRa [$\epsilon = 1\%$] & 0.0\% &0.0\% &0.0\% &0.0\% &0.0\% &0.0\% &0.0\% \\
		BlRa [$\epsilon = 10\%$] & 0.0\% &0.0\% &0.0\% &0.0\% &0.0\% &0.0\% &0.0\% \\
		BlRa [$\epsilon = 25\%$] & 0.0\% &0.0\% &0.0\% &0.0\% &0.0\% &0.0\% &0.0\% \\
		\bottomrule
	\end{tabular}
	\caption{GSet - redundancy ratios}
	\label{tab:gset_redundancy_ratios}
\end{table*}

\subsection{Bloom + Bucketing vs Bloom + Bucketing + Rateless}\label{appendix:bloom-bucketing-rateless}
\begin{figure*}
    \centering
    \includegraphics[width=\textwidth]{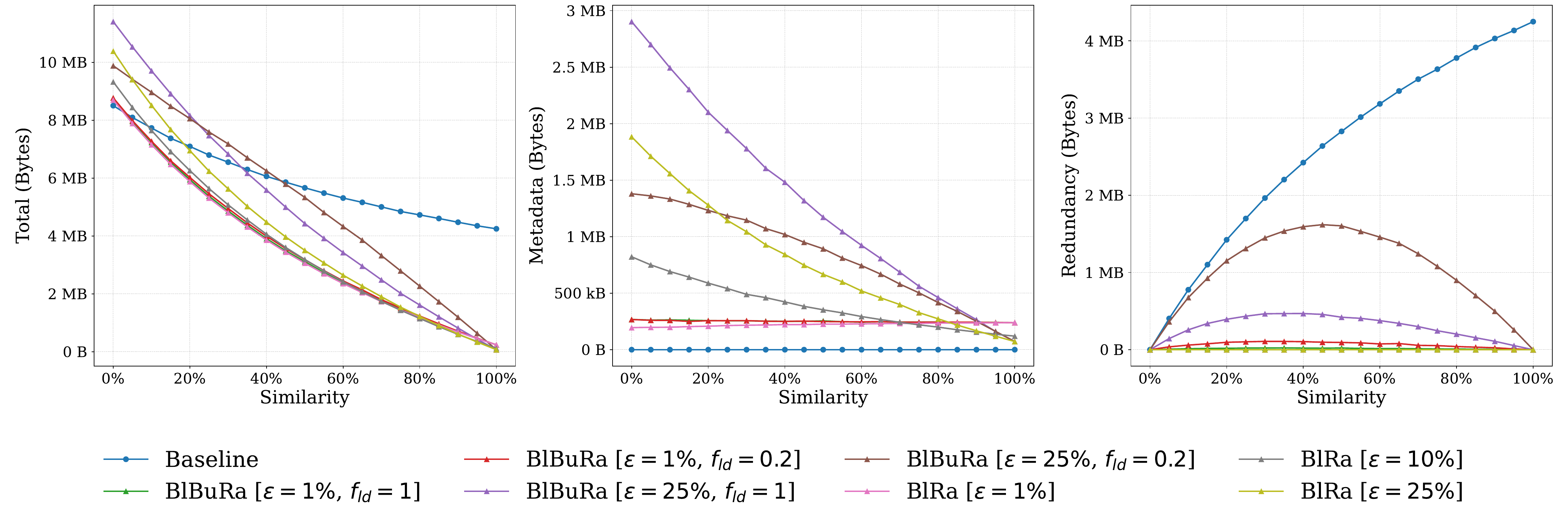}
    \caption{Transmission analysis w.r.t. similarity between a pair of GSets - Bloom + Bucketing vs Bloom + Bucketing + Rateless}
    \label{fig:bloom-bucketing-transmitted}
\end{figure*}

\subsection{High Similarity}\label{appendix:high-similarity}

\end{document}